\documentstyle[aps,prl,psfig,floats]{revtex}

\begin{document}

\twocolumn[\hsize\textwidth\columnwidth\hsize\csname@twocolumnfalse\endcsname

\vskip -0.5truecm
\title{Structural and electronic properties of a wide-gap quaternary alloy:\\
${\rm Zn_xMg_{1-x}S_ySe_{1-y}}$}

\author{A.M. Saitta,$^{1,}$\cite{present,e-mail} 
S. de Gironcoli,$^1$\cite{e-mail} and S. Baroni$^{1,2}$\cite{e-mail} }

\address{$^1$INFM -- Istituto Nazionale per la Fisica della Materia
and \\ SISSA -- Scuola Internazionale Superiore di Studi Avanzati,
Via Beirut 2-4, I-34014 Trieste, Italy \\ $^2$CECAM -- Centre
Europ\'een de Calcul Atomique et Mol\'eculaire, ENSL, 46~All\'ee
d'Italie, 69364 Lyon Cedex 07, France} 

\date{\today}
\maketitle

\begin{abstract}
The structural properties of the ${\rm Zn_xMg_{1-x}S_ySe_{1-y}}$ solid 
solutions are determined by a combination of the {\it computational
alchemy} and the {\it cluster expansion} methods with Monte Carlo
simulations. We determine the phase diagram of the alloy
and show that the homogeneous phase is characterized by a large
amount of short-range order occurring among first-nearest neighbors.
Electronic-structure calculations performed using the {\it special
quasi-random structures} approach indicate that the energy gap of
the alloy is rather sensitive to this short-range order.
\end{abstract}

\vspace{-0.5truecm}
] 
\narrowtext

Wide gap semiconductors have recently attracted an enormous
technological interest~\cite{II_VI_Review} 
both because of their potential use in devices
capable of operating at high power level and high temperature, and
because of the need for optical materials active in the blue-green
spectral range. ZnSe-based technology will be used for operation in
this spectral range provided that current device lifetime problems are
overcome. One major goal of materials engineering for opto-electronic
applications is the ability to tune {\it independently} the band gap,
$E_g$---in order to obtain the desired optical properties---and the
lattice parameter, $a_0$, of the material---in order to be able to grow
it on a given substrate. Unfortunately, in most III-V and II-VI
alloys, the additional degree of freedom provided by alloying both the
cationic and the anionic sub-lattices cannot be effectively exploited
because $E_g$ and $a_0$ are roughly inversely proportional to one
another for any values of the cationic and anionic compositions,
$(x,y)$. From this point of view, ${\rm Zn_xMg_{1-x}S_ySe_{1-y}}$
alloys play a special role, in that the lattice
parameter and optical gap can be varied fairly independently as
functions of $(x,y)$ \cite{ZnMgSSe}.

In spite of this big interest, many technical difficulties still
hinder a precise experimental characterization of these materials, so
that their equilibrium structural and optical properties are basically
unknown. In this Letter we report on the first application of
state-of-the-art electronic-structure techniques to the determination
of the structural and optical properties of a quaternary (double
binary) semiconductor alloy at thermodynamic equilibrium, and present
results in the specific case of ${\rm Zn_xMg_{1-x}S_ySe_{1-y}}$. Even
though MBE-grown materials are fabricated in highly non-equilibrium
conditions, understanding their equilibrium properties is preliminary
to any further investigations and provides the physical limits to the
tunability of the alloy properties.

The first goal of this Letter is to determine the thermodynamic
stability of the {\rm (Zn,Mg)(S,Se)} solid solution with respect to
segregation into its constituents and/or to the formation of ordered
structures. Secondly, we will analyze how the fundamental gap depends
on compositions and the role that short-range order plays in the
electronic properties of this material.

The thermodynamic properties of ${\rm Zn_xMg_{1-x}S_ySe_{1-y}}$ are
studied by mapping the alloy onto a (double) lattice-gas
model~\cite{DdF,CW} that is solved by standard Monte Carlo techniques.
To this end, an Ising-like variable, ${\rm \{ \sigma_{{\bf R}s}\}}$,
is first attached to the $s$-th atom in the elementary cell located at
${\rm \bf R}$, and it is assumed to take the values $\pm 1$ according
to the type of atom occupying that lattice site. 
The energy of the alloy is then expanded in terms of the $\sigma$'s
including terms up to three-spin interactions:
\begin{eqnarray}\nonumber &E&\bigl (\{ \sigma \} \bigr ) = \hfill E_0+
\sum\limits_{{\bf R}s} K_s \sigma _{{\bf R}s}+\frac 12 \sum \limits_{
{\bf RR^{\prime }} \atop \scriptstyle ss^\prime}J_{ss^\prime} \left(
{\bf R-R^{\prime }}\right) \,\sigma_{{\bf R}s} \sigma_{{\bf R^{\prime
}}s^\prime} + \\
&+&\frac 1{6}\sum\limits_{{\bf RR}^{\prime }{\bf R^{\prime
\prime }} \atop\scriptstyle ss^\prime s^{\prime\prime}} L_{ss^\prime
s^{\prime\prime}} \left({\bf R-R^{\prime \prime },R^{\prime
}-R^{\prime \prime }}\right) \sigma _{{\bf R}s}\sigma _{{\bf R^{\prime
}}s^\prime} \sigma _{{\bf R^{\prime \prime}}s^{\prime\prime}}. \end{eqnarray} 
\narrowtext
\noindent The validity of the truncation is then established {\it a
posteriori}. The two- and three-body interaction constants, $J$
and $L$, can in principle be obtained from both the {\it cluster
expansion} (CE) \cite{Zunger} or {\it computational alchemy} (CA)
\cite{Alchemy} approaches. The explicit inclusion of local
distortions from the ideal lattice positions---due to the different
atomic sizes---renormalizes the two-body interaction, making
it long-ranged \cite{Alchemy} and hence hardly obtainable by the CE
method. In principle, the three-body interactions, $L$, could also be
computed within CA by making use of the so called {\em 2n+1 theorem}
\cite{2n1}. Assuming that the $L$'s are short-ranged, however, they
are more conveniently obtained by the CE method. To this end, we first
calculate the $J$'s up to 9-th nearest neighbors from CA by using
second-order density-functional perturbation theory
\cite{technicalities}. We then assume that the only non-vanishing
$L$'s are those which connect atoms that are first- or, at most,
second-nearest neighbors. Because of symmetry, there are only six
such 3-body interaction constants. In the zincblende structure, 
there are two inequivalent triplets of second-nearest neighbors which differ
according to whether or not they have a common first-nearest neighbor.
We follow the common practice of neglecting such difference \cite{Zunger},
and we keep 4 inequivalent 3-body interactions. In order to determine the 
latter, we have calculated the total energies of a set of 28 pseudo-binary
ordered structures by using self-consistent density-functional theory
(DFT) within the local-density approximation (LDA)
\cite{technicalities}, and by allowing the lattice to relax until the
forces acting on individual atoms vanish. We have then fitted the
cubic term of Eq. (1) to the differences between the energies so
calculated and the predictions of Eq. (1), when truncated to second
order. The quality of the fit so obtained is finally checked against
39 additional quaternary structures. The resulting mean-square error
is $ \lesssim 0.8~ \rm meV$ per atom pair, while the maximum error is
$ 2.1~ \rm meV$. Two- and three-body constants have been calculated
for 5 and 3 volumes in the range $[V_{\rm ZnS},V_{\rm MgSe}]$
respectively, and interpolated in-between.

Finite-temperature properties have been determined by lattice-gas
Monte Carlo simulations, using supercells of 1024 atoms, at constant
temperature, pressure, and chemical potentials. To this end, we
attempt two different types of Monte Carlo moves: the reversal
(flip) of each spin and a variation of the volume of the
supercell. The trial move is then accepted with the Metropolis
probability. Measures were taken over $\approx 10^3$ correlation times
after thermal equilibrium was reached.

The free energy surface of the system is extracted from the
determination of the compositions, $x$ and $y$, as functions of the
chemical potentials, $\mu_x$ and $\mu_y$. The regions of {\em
spinodal} (local) stability correspond to those concentrations for
which the Hessian of the free energy with respect to the 
concentrations is 
positive definite. The thermodynamically stable concentrations
(binodal regions), and thus the miscibility gaps, are determined with
a generalization of the common-tangent Maxwell construction usually
adopted for binary mixtures: the globally stable points are those 
which are locally stable and whose tangent planes do not intersect the 
free energy surface at any other points in the square of concentrations.

\begin{figure}
\centerline{\psfig{figure=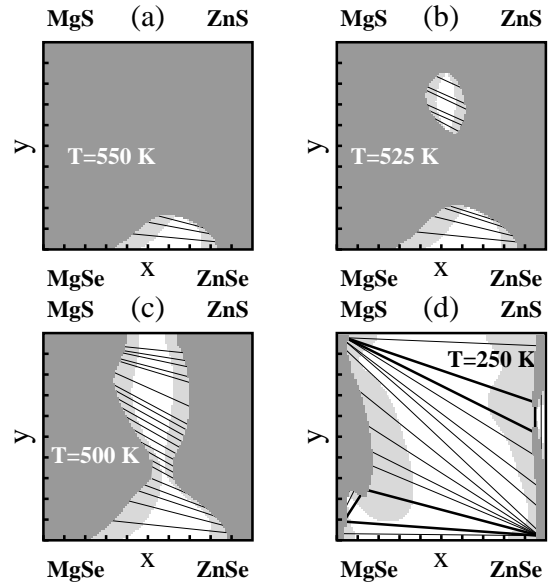,width=8.5cm}}
\caption{Phase diagram of the alloy at four different temperatures.
The dark gray (binodal) regions are thermodynamically stable.
The light gray (spinodal) regions are only locally stable (metastable). 
The white regions are completely unstable.
The segments inside the miscibility gap connect the stable phases
identified by the common-tangent-plane construction (see text). 
In panels $a-c$ only separation into two phases occur. In panel $d$
two cases where the products of separation are three distinct phases
are indicated by thicker lines.
}
\label{phasediagram}
\vspace{-0.3truecm}
\end{figure}

In Fig.~\ref{phasediagram} we display a cut of the phase diagram of
the quaternary alloy at four different temperatures. At typical MBE
growth temperatures ($\approx 600~K$) and above the alloy is predicted
to be completely miscible. For $T=550~K$ (Fig.~\ref{phasediagram}a),
the disordered alloy is stable at all compositions but for a small
island close to the ${\rm Zn_{x}Mg_{1-x}Se}$ pseudo-binary alloy
centered around $x\approx 0.6$, whose computed critical temperature is
of about $610~K$. Decreasing the temperature
(Fig.~\ref{phasediagram}b) the size of the island increases and a
forbidden region appears inside the square of compositions, close to
the mid point $(x,y)=({\frac 12},{\frac 12})$. Cooling down to
$T=500~K$ (Fig.~\ref{phasediagram}c), the ${\rm Mg}$-rich and the
${\rm Zn}$-rich regions become separated by a ``corridor'' in which the
disordered alloy is not locally stable. The concentrations of the
phases in which the alloy segregates in the miscibility gap are given
by the contact points of the tangent plane. Segregation may thus
result in the separation into two or three phases, according to the number of
contact points. In the present case, the critical temperatures of the
pseudo-binary alloys with cationic disorder, ${\rm Zn_xMg_{1-x}Se}$ and
${\rm Zn_xMg_{1-x}S}$, are much larger ($T_c=613~K$ and $511~K$
respectively) than those of the pseudo-binary alloys with anionic
disorder, ${\rm ZnS_ySe_{1-y}}$ ($T_c=254~K$) and ${\rm
MgS_ySe_{1-y}}$ ($T_c=243~K$); therefore, above room temperature the
alloy segregates into two phases: a ${\rm Zn}$-rich phase and a ${\rm
Mg}$-rich one. This behavior is due to the larger chemical difference
between cations than between anions. At temperatures of the order of the 
anionic-alloy critical temperatures and below
($T\lesssim 250~K$), the tendency to segregation also involves the
anionic sub-lattice, and separation into three phases may occur
(Fig.~\ref{phasediagram}d).

Our results show that in (Zn,Mg)(S,Se) the onset of segregation occurs
at a critical temperature that is in the range of typical MBE growth
temperatures, and thus much lower than in other II-VI quaternary
alloys~\cite{Kisker}. This result is in agreement with an analysis of
experimental data~\cite{EXAFS}, based on {\it delta-lattice-parameter}
models~\cite{DLP}, that locate $T_c$ between $525~K$ and
$625~K$. According to the only experimental report~\cite{3M1} we are
aware of, a forbidden region of compositions has been observed, at
room temperature, inside the predicted miscibility gap.

\begin{table}
\centerline{$a_0(x,y)=a_{\rm MgSe}+Ax+By+Cxy+Dx^2+Ey^2$}
\begin{tabular}{lcccccc}
 & $a_{\rm MgSe}$ & A & B & C & D & E \\
\hline
LDA & 11.326 & -0.789 & -0.539 & 0.058 & 0.027 & 0.016 \\
Corrected LDA & 11.130 & -0.443 & -0.523 & 0.018 & 0.027 & 0.016 \\
\end{tabular}
\caption{Quadratic fit coefficients for LDA (upper row) and
``corrected'' LDA (lower row) lattice constant of the alloy. 
The maximum error of the
fit is of 0.05\%, and the mean square error of 0.01\%.}
\label{t:Struct}
\vspace{-0.2truecm}
\end{table}
The structural properties of the alloy can be extracted from our
Monte Carlo simulations. The equilibrium lattice constant has a strong
linear dependence upon compositions, thus following Vegard's
law. Slight deviations from this law are extracted by a fit of the
lattice parameter of the alloy with a quadratic polynomial in $x$ and
$y$ (see Table~\ref{t:Struct}). A direct comparison with experimental
data may be done by correcting for the LDA error on the equilibrium
lattice constants of the pure constituents of the alloy ($\pm 1\div
2\%$), which is accounted for by a bilinear interpolation and added to
the quadratic polynomial obtained above. The corrected coefficients of
polynomial fit are also reported in Table~\ref{t:Struct}. The
agreement with experimental data~\cite{ZnMgSSe} is good in the ${\rm
Zn}$-rich region, especially in proximity of pure ${\rm ZnSe}$, while
it is apparently poor in the ${\rm Mg}$-rich part of the square of the
compositions, where however the instability of the system makes it
very difficult to measure the concentrations with the same precision,
and the experimental data are in fact very scarce.

\begin{figure}
\psfig{figure=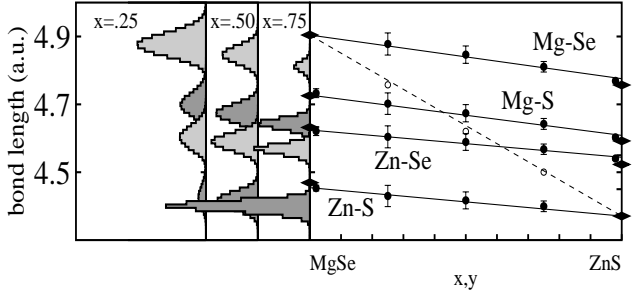,width=8.5cm}
\caption{Left panel: bond-length distributions along the ${\rm
ZnS-MgSe}$ diagonal in the square of the compositions. Right panel:
dependence of maxima of the peak upon the concentrations. Diamonds
correspond to pure-material bond-lengths. Empty dots refer to the
alloy lattice constant and the dashed line is a linear fit.}
\label{bonds}
\vspace{-0.5truecm}
\end{figure}
As it is the case in other tetrahedrally bonded solid solutions, the
Vegard's law does not result from a linear dependence of individual
bond lengths upon compositions, but rather from a subtle topological
compensation of bond lengths which stay in fact rather close to the
values they would have in pure compounds \cite{EXAFS}. This behavior
is displayed in Fig.~\ref{bonds}, where we show the bond lengths as
obtained at T=$800~K$ for the {\rm (Zn,Mg)(S,Se)} alloy along the
MgSe-ZnS diagonal of the square of concentrations (along the other
diagonal the lattice parameter is almost constant and matched to that
of GaAs). At any concentrations, while the lattice parameter of the
alloy varies almost linearly by more than 12~\%, the largest
deviations between the bond lengths and their pure-compound values are
smaller than 3\%.

The typical substrate on which such alloys are grown is ${\rm GaAs}$,
whose lattice constant is $a_0=10.68~{\rm a.u.}$, close to the
equilibrium lattice parameters of ${\rm ZnSe}$ and ${\rm MgS}$.
Therefore, the technological relevant concentrations of the quaternary
alloy are those located along the ZnSe-MgS diagonal of the square of
the compositions. For these solid solutions, the Mg-Se and Zn-S bonds
are consequently subject to a larger elastic strain with respect to
the other bonds; the preferential formation of MgS and ZnSe clusters
is thus expected to be energetically favored.  In order to clarify
this issue we have calculated the two-body correlations of the ${\rm
Zn_{\frac 12}Mg_{\frac 12}S_{\frac 12}Se_{\frac 12}}$ alloy at
temperatures above the miscibility gap. The correlation function is
defined as: $ C_{s s^\prime}({\bf
R})=\langle \sigma_{s\bf 0} \sigma_{s^\prime{\bf R}}\rangle -\langle 
\sigma_{s}\rangle \langle \sigma_{s^\prime}\rangle$.
Therefore, $C_{s s^\prime}({\bf R})=0$ in the perfectly
random alloy, $C_{s s^\prime}({\bf R})>0$ for ZnSe and MgS
clusterizations, and $C_{s s^\prime}({\bf R})<0$ for ZnS and MgSe
clusterizations. In correspondence of the first-nearest-neighbor
shell, $C_{s s^\prime}({\bf R})$ has a very pronounced positive peak,
thus indicating a strong tendency to form ZnSe and MgS clusters. This
tendency is confirmed by the positive value of the cation-cation or
anion-anion correlations in the second-nearest-neighbor shell, which
are however weaker. Correlations practically die beyond the second
shell of neighbors, and are therefore characteristic of a very
pronounced short-range order (SRO). 
\begin{figure}
\psfig{figure=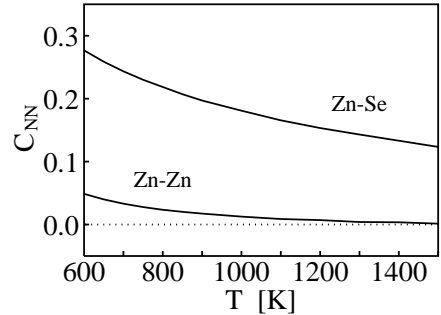,width=6.5cm}
\caption{First- and second-nearest-neighbors two-body correlations
of the ${\rm Zn_{\frac 12}Mg_{\frac 12}S_{\frac 12}Se_{\frac 12}}$ alloy
as function of temperature.}
\label{correlations}
\vspace{-0.3truecm}
\end{figure}
In Fig.~\ref{correlations} we
display the value of the first- and second-shell correlation peaks for
a range of temperatures above $T_c$. The nearest-neighbor correlation
peak is still significantly large at very high temperature, thus
indicating that short-range-order is present even close to the melting
temperatures (above $T\approx 1700~K$): the system can never be
described as a perfectly random alloy.

The second goal of this work is to study the influence of the
structural properties of the alloy at the atomistic level on the
electronic and optical properties. The band structure of a disordered
material is strongly affected by its local environment that is 
poorly approximated by effective-medium approaches~\cite{VCA,CPA}. A
detailed account of the microscopic structure of the alloy is even
more important in the present case where SRO
effects are expected to play a significant role. A ``direct'' DFT-LDA
calculation should be in principle performed by using very large
supercells in order to cope with compositional disorder. Here
we adopt the {\em special quasi-random structures} (SQS)
approach~\cite{SQS}, suitably generalized to account for
SRO effects~\cite{Mader},
double-sublattice disorder, and arbitrary compositions.
We have considered three different concentrations along the line of
lattice matching to GaAs, and chosen among the discrete set of
compositions compatible with 64-atom SQS's and close to the ${\rm
ZnSe}$-rich region: $(x,y)=({\frac 12},{\frac 12});({\frac 34},{\frac
18});({\frac {27}{32}},0)$. For these concentrations, SQS's have been
obtained by a simulated-annealing procedure aimed at modeling the pair
correlation functions at short range---as obtained from Monte Carlo
simulations performed at $T=550~K$ and resulting in very faithful
pair correlations up to fourth-nearest-neighbors.  We
have explicitly verified that the calculated electronic properties of
the alloy are rather insensitive to correlations beyond this order of
neighbors.

\begin{figure}
\centerline{\psfig{figure=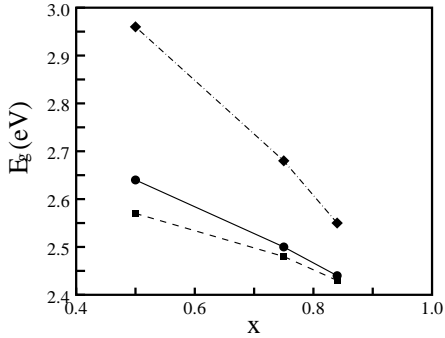,width=6.5cm}}
\caption{
Energy gap of the ${\rm (Zn,Mg)(S,Se)}$ quaternary alloy 
lattice-matched to ${\rm GaAs}$ as function
of the $x$ composition (bottom scale) and $y$ composition (top scale). 
The dotdashed line refers to virtual crystal calculations,
the dashed one to perfectly-random-alloy results, 
and the solid line to the short-range-ordered alloy band gaps.
}
\label{gaptot}
\vspace{-0.3truecm}
\end{figure}
It is interesting to compare the results obtained using different
levels of approximations to substitional disorder. In
Fig.~\ref{gaptot} we display the energy gap as a function of the
cationic composition along the ${\rm GaAs}$-matching line of the 
compositions plane, as obtained from calculations performed {\it i)}
on the appropriate virtual crystal, {\it ii)} on supercells describing
the perfectly random alloy, and {\it iii)} on SQS's that reproduce the
SRO correlations as explained above. We see that virtual-crystal is a
bad approximation of the real system, while the effects of SRO show up
in a slight but non-negligible opening of the fundamental band gap. 
Our results, once bilinearly corrected for the DFT error in the pure materials,
agree very well (within 40 meV), with freshly appeared experimental estimates
\cite{Egap_exp}.
An analysis of the momentum- and position-projected density of states
shows that the fundamental gap
is direct and occurs at the $\Gamma$ point of the Brillouin zone
for any concentrations. The top-valence band states are mainly
${\rm Se}$ states, while the bottom state of the conduction band is rather
delocalized on the different atomic species and it is found to be the
most sensitive to the occurrence of short-range-order.

We wish to thank A.\ Franciosi and S.\ Rubini for prompting our interest
in this problem and for many fruitful discussions. A more complete account
of this work can be found in the PhD thesis of one of us (AMS) which is 
available on the www\cite{PhD}. This work has been done in part within the 
{\it Iniziativa Trasversale Calcolo Parallelo} of the INFM.

\end{document}